\documentclass[12pt]{article}
\usepackage[T2A]{fontenc}
\usepackage[english]{babel}

\usepackage{latexsym}
\usepackage{amssymb}
\usepackage{amsmath}

\def\proof{\medbreak\noindent{\bf Proof}}

\def\theorem #1. #2\par{\medbreak
  \noindent{\tt {\bf Theorem #1.}\enspace}{\sl#2\par}%
  \ifdim\lastskip<\medskipamount \removelastskip\penalty55\medskip\fi}

\def\lemma #1. #2\par{\medbreak
  \noindent{\tt {\bf Lemma #1.}\enspace}{\sl#2\par}%
  \ifdim\lastskip<\medskipamount \removelastskip\penalty55\medskip\fi}

\def\{{\lbrace}
\def\}{\rbrace}

\def\Wcl{W{\cal C}\!\ell}
\def\wcl{w{\cal C}\!\ell}

\def\cl{{\cal C}\!\ell}

\def\R{{\Bbb R}}

\def\be{\begin{equation}}
\def\ee{\end{equation}}

\newcommand{\fin}{\hbox{$\bullet$}\medskip}

\begin{document}
\title{A classification of Lie algebras of pseudounitary groups in the techniques of Clifford algebras}

\author{Shirokov D.S.}

\begin{abstract}
In this paper we present new formulas, which represent commutators
and anticommutators of Clifford algebra elements as sums of elements
of different ranks. Using these formulas we consider subalgebras of
Lie algebras of pseudounitary groups. Our main techniques are
Clifford algebras. We have find 12 types of subalgebras of Lie
algebras of pseudounitary groups.
\end{abstract}

\maketitle

\bigskip
In this paper we revise and further develop some results of
\cite{Marchuk:Shirokov}. Namely, we make more precise the formulas
for commutators and anticommutators of Clifford algebra elements of
fixed ranks (Theorem 1 and 2).

We investigate Lie algebras of pseudounitary groups using the
techniques of Clifford algebra. We present 12 types of subalgebras
of Lie algebras of pseudounitary groups (Theorem 3).

\section{Formulas for commutators and anticommutators of Clifford algebra elements}

Let $p, q$ be nonnegative integer numbers and $p+q=n$, $n\geq1$. Consider the complex Clifford algebra $\cl(p,q)$ \cite{Marchuk:Shirokov}. Let $e$ be the identity element and let $e^a$, $a=1,\ldots,n$ be generators of the Clifford algebra $\cl(p,q)$,
$$
e^a e^b+e^b e^a=2\eta^{ab}e,
$$
where $\eta=||\eta^{ab}||$ is the diagonal matrix with $p$ pieces of $+1$ and $q$ pieces of $-1$ on the diagonal. Elements
$$
e^{a_1\ldots a_k}=e^{a_1}\ldots e^{a_k},\quad a_1<\ldots<a_k,\,k=1,\ldots,n,
$$
together with the identity element $e$, form a basis of the Clifford
algebra. The number of basis elements is equal to $2^n$. We denote
by $\cl_k(p,q)$ the vector spaces that span over the basis elements
$e^{a_1\ldots a_k}$. Elements of $\cl_k(p,q)$ are said to be
elements of rank $k$.

The construction of Clifford algebra $\cl(p,q)$ is discussed in details in \cite{Marchuk:Shirokov}. The following theorem makes more precise the statement of Theorem 7 of \cite{Marchuk:Shirokov}.

\begin{theorem}1.
Let $\stackrel{k}{U}, \stackrel{l}{V}, \stackrel{r}{W}$ be Clifford
algebra elements of ranks $k, l$, and $r$ respectively. Then, for
all integer nonnegative numbers $n\geq k\geq l\geq 0$, the following
formulas are valid.

1) If $n\geq k+l$, then for $l\neq 0$
\begin{equation}
[\stackrel{k}{U},\stackrel{l}{V}]=\left\lbrace
\begin{array}{ll}
\stackrel{k-l+2}{W}+\stackrel{k-l+6}{W}+\ldots+\stackrel{k+l-2}{W}, & \mbox{\rm $l$ - even;}\\
\stackrel{k-l}{W}+\stackrel{k-l+4}{W}+\ldots+\stackrel{k+l-2}{W}, & \mbox{\rm $k$ - even, $l$ - odd;}\\
\stackrel{k-l+2}{W}+\stackrel{k-l+6}{W}+\ldots+\stackrel{k+l}{W}, &
\mbox{\rm $k,l$ - odd}
\end{array}
\right.
\label{1}
\end{equation}
and
\begin{equation}
[\stackrel{k}{U},\stackrel{0}{V}]=0. \label{2}
\end{equation}
2) If $k+l\geq n$, then for $k\neq n$
\begin{equation}
[\stackrel{k}{U},\stackrel{l}{V}]=\left\lbrace
\begin{array}{ll}
\stackrel{k-l}{W}+\stackrel{k-l+4}{W}+\ldots+\stackrel{2n-k-l}{W}, &
\mbox{\rm $n$ - even,
$k$ - even, $l$ - odd;}\\ \\
\stackrel{k-l}{W}+\stackrel{k-l+4}{W}+\ldots+\stackrel{2n-k-l-2}{W}, & \mbox{\rm $n$ - odd, $k$ - even, $l$ - odd;}\\ \\
\stackrel{k-l+2}{W}+\stackrel{k-l+6}{W}+\ldots+\stackrel{2n-k-l}{W},
&
\parbox{
.5\linewidth}{$n$ - even, $k$ - odd or\\ $n$ - odd, $k$ - even, $l$ - even;}\\ \\
\stackrel{k-l+2}{W}+\stackrel{k-l+6}{W}+\ldots+\stackrel{2n-k-l-2}{W},
&
\parbox{ .5\linewidth}{$n$ - odd, $k$ - odd or\\ $n$ - even, $k$ -
even, $l$ - even}
\end{array}
\right.
\label{3}
\end{equation}
and
\begin{equation}
[\stackrel{n}{U},\stackrel{l}{V}]=\left\lbrace
\begin{array}{ll}
0, & \parbox{.4\linewidth}{$n$ - even, $l$ - even or\\ $n$ - odd;}\\ \\
\stackrel{n-l}{W}, & \mbox{\rm $n$ - even, $l$ - odd.}
\end{array}
\right.
\label{4}
\end{equation}
\end{theorem}

\proof. \,
Note that any Clifford algebra element is a linear combination of basis elements. Let's prove our theorem for basis elements of the Clifford algebra.

Let us take
\begin{equation}
e^{a_1\ldots a_k} e^{b_1\ldots b_l} \in\cl_{k+l-2s}(p,q),
\end{equation}
where $s$ is the number of coincident indices in ordered multi-indices
$a_1\,\ldots\,a_k$ and $b_1\,\ldots\,b_l$. Here $\cl_{k+l-2s}(p,q)$ can be considered as vector space that spans over the elements $e^{a_1\ldots a_{k+l-2s}}$.
Since $e^a e^b+e^b e^a=2\eta^{ab}e$, it follows that
\begin{eqnarray*}
[e^{a_1\ldots a_k}, e^{b_1\ldots b_l}] & = &
(1-(-1)^{kl-s})e^{a_1\ldots a_k}e^{b_1\ldots b_l}.
\end{eqnarray*}

Finally, we obtain
$$
[e^{a_1\ldots a_k}, e^{b_1\ldots b_l}]=
\left\lbrace
\begin{array}{ll}
\stackrel{k+l-2s}{W}, & \mbox{if $kl-s$ is odd},\\
0,  & \mbox{if $kl-s$ is even}.
\end{array}
\right.
$$
For $n\geq k+l$ we have $0 \leq s \leq l$. And for $k+l\geq n$ number $s$ takes values from $k+l-n$ to $l$. Considering all possible values of $s$ and taking into account evenness of $kl-s$, we complete the proof of Theorem 1.
\fin

\begin{theorem}2.
Let $\stackrel{k}{U}, \stackrel{l}{V}, \stackrel{r}{W}$ be Clifford
algebra elements of the ranks $k, l$, and $r$ respectively. Then,
for all integer nonnegative numbers $n\geq k\geq l\geq 0$, the
following formulas are valid.

1) If $n\geq k+l$, then for $l\neq 0$
$$
\{\stackrel{k}{U},\stackrel{l}{V}\}=\left\lbrace
\begin{array}{ll}
\stackrel{k-l}{W}+\stackrel{k-l+4}{W}+\ldots+\stackrel{k+l}{W}, & \mbox{\rm $l$ - even;}\\
\stackrel{k-l+2}{W}+\stackrel{k-l+6}{W}+\ldots+\stackrel{k+l}{W}, & \mbox{\rm $k$ - even, $l$ - odd;}\\
\stackrel{k-l}{W}+\stackrel{k-l+4}{W}+\ldots+\stackrel{k+l-2}{W}, &
\mbox{\rm $k,l$ - odd}
\end{array}
\right.
$$
and
$$\{\stackrel{k}{U},\stackrel{0}{V}\}=\stackrel{k}{W}.$$

2) If $k+l\geq n$, then for $k\neq n$
$$
\{\stackrel{k}{U},\stackrel{l}{V}\}=\left\lbrace
\begin{array}{ll}
\stackrel{k-l+2}{W}+\stackrel{k-l+6}{W}+\ldots+\stackrel{2n-k-l}{W},
& \mbox{\rm $n$ - odd,
$k$ - even, $l$ - odd;}\\ \\
\stackrel{k-l+2}{W}+\stackrel{k-l+6}{W}+\ldots+\stackrel{2n-k-l-2}{W}, & \mbox{\rm $n$ - even, $k$ - even, $l$ - odd;}\\ \\
\stackrel{k-l}{W}+\stackrel{k-l+4}{W}+\ldots+\stackrel{2n-k-l}{W}, &
\parbox{.5\linewidth}
{$n$ - odd, $k$ - odd or\\ $n$ - even, $k$ - even, $l$ - even;}\\ \\
\stackrel{k-l}{W}+\stackrel{k-l+4}{W}+\ldots+\stackrel{2n-k-l-2}{W},
&
\parbox{.5\linewidth} {$n$ - even, $k$ - odd or \\ $n$ - odd, $k$ -
even, $l$ - even}
\end{array}
\right.
$$
and
$$
\{\stackrel{n}{U},\stackrel{l}{V}\}=\left\lbrace
\begin{array}{ll}
0, & \mbox{\rm $n$ - even, $l$ - odd;}\\ \\
\stackrel{n-l}{W}, & \parbox{.3\linewidth}{$n$ - odd or\\ $n$ -
even, $l$ - even.}
\end{array}
\right.
$$
\end{theorem}

\proof. \,
The proof is analogous to the proof of Theorem 1.
\fin

Let's write down some special cases of formulas for commutators and anticommutators of Clifford algebra elements from Theorem 1 and Theorem 2.

If ranks of two Clifford algebra elements are equal (k=l), then
$$
[\stackrel{k}{U},\stackrel{k}{V}]=\left\lbrace
\begin{array}{ll}
\stackrel{2}{W}+\stackrel{6}{W}+\ldots+\stackrel{2k}{W}, & \mbox{\rm k - odd and $n\geq 2k$;}\\
\stackrel{2}{W}+\stackrel{6}{W}+\ldots+\stackrel{2k-2}{W}, & \mbox{\rm k - even and $n\geq 2k$;}\\
\stackrel{2}{W}+\stackrel{6}{W}+\ldots+\stackrel{2n-2k}{W}, & \mbox{\rm $2k\geq n$ and n,k are of different evenness;}\\
\stackrel{2}{W}+\stackrel{6}{W}+\ldots+\stackrel{2n-2k-2}{W}, & \mbox{\rm $2k\geq n$ and n,k are of same evenness;}\\
0, & \mbox{\rm $k=n$ or $k=0$.}
\end{array}
\right.
$$

$$
\{\stackrel{k}{U},\stackrel{k}{V}\}=\left\lbrace
\begin{array}{ll}
\stackrel{0}{W}+\stackrel{4}{W}+\ldots+\stackrel{2k-2}{W}, & \mbox{\rm k - odd and $n\geq 2k$;}\\
\stackrel{0}{W}+\stackrel{4}{W}+\ldots+\stackrel{2k}{W}, & \mbox{\rm k - even and $n\geq 2k$;}\\
\stackrel{0}{W}+\stackrel{4}{W}+\ldots+\stackrel{2n-2k-2}{W}, & \mbox{\rm $2k\geq n$ and n,k are of different evenness;}\\
\stackrel{0}{W}+\stackrel{4}{W}+\ldots+\stackrel{2n-2k}{W}, & \mbox{\rm $2k\geq n$ and n,k are of same evenness;}\\
0, & \mbox{\rm $k=n$ or $k=0$.}
\end{array}
\right.
$$

If one rank is fixed, then
\begin{flalign*}
[\stackrel{a}{U},\stackrel{1}{V}]&=\left\lbrace
\begin{array}{ll}
\stackrel{a-1}{W}, & \mbox{\rm a - even;}\\
\stackrel{a+1}{W}, & \mbox{\rm a - odd, $a\neq n$;}\\
0, & \mbox{\rm a - odd, $a=n$.}
\end{array}
\right. &\qquad
\lbrace\stackrel{a}{U},\stackrel{1}{V}\rbrace&=\left\lbrace
\begin{array}{ll}
\stackrel{a-1}{W}, & \mbox{\rm a - odd;}\\
\stackrel{a+1}{W}, & \mbox{\rm a - even, $a\neq n$;}\\
0, & \mbox{\rm a - even, $a=n$.}
\end{array}
\right.
\\
[\stackrel{a}{U},\stackrel{2}{V}]&=\left\lbrace
\begin{array}{ll}
\stackrel{a}{W}, & \mbox{\rm $a\neq n$;}\\
0, & \mbox{\rm $a=n$.}
\end{array}
\right. &\qquad
\lbrace\stackrel{a}{U},\stackrel{2}{V}\rbrace&=\left\lbrace
\begin{array}{ll}
\stackrel{a-2}{W}+\stackrel{a+2}{W}, & \mbox{\rm $a\neq n, n-1$;}\\
\stackrel{a-2}{W}, & \mbox{\rm $a=n, n-1$.}
\end{array}
\right.
\\
[\stackrel{a}{U},\stackrel{3}{V}]&=\left\lbrace
\begin{array}{ll}
\stackrel{a-3}{W}+\stackrel{a+1}{W}, & \parbox{
.2\linewidth}{a - even,\\ $a\leq n-2$;}\\ \\
\stackrel{a-3}{W}, & \parbox{
.2\linewidth}{ a - even,\\ $a=n-1, n$;}\\ \\
\stackrel{a-1}{W}+\stackrel{a+3}{W}, & \parbox{
.2\linewidth}{a - odd,\\ $a\leq n-3$;}\\ \\
\stackrel{a-1}{W}, & \parbox{
.2\linewidth}{a - odd,\\ $a=n-2, n-1$;}\\ \\
0, & \parbox{.2\linewidth}{a - odd,\\ $a=n$.}
\end{array}
\right. &\qquad
\lbrace\stackrel{a}{U},\stackrel{3}{V}\rbrace&=\left\lbrace
\begin{array}{ll}
\stackrel{a-3}{W}+\stackrel{a+1}{W}, & \parbox{
.2\linewidth}{a - odd,\\ $a\leq n-2$;}\\ \\
\stackrel{a-3}{W}, & \parbox{
.2\linewidth}{a - odd,\\ $a=n-1, n$;}\\ \\
\stackrel{a-1}{W}+\stackrel{a+3}{W}, & \parbox{
.2\linewidth}{a - even,\\ $a\leq n-3$;}\\ \\
\stackrel{a-1}{W}, & \parbox{
.2\linewidth}{a - even,\\ $a=n-2, n-1$;}\\ \\
0, & \mbox{\rm a - even, $a=n$.}
\end{array}
\right.
\\
[\stackrel{a}{U},\stackrel{4}{V}]&=\left\lbrace
\begin{array}{ll}
\stackrel{a-2}{W}+\stackrel{a+2}{W}, & \parbox{
.2\linewidth}{$a\leq n-3$;}\\ \\
\stackrel{a-2}{W}, & \parbox{
.2\linewidth}{$a=n-2,\\ n-1$;}\\ \\
0, & \mbox{\rm $a=n$.}
\end{array}
\right. &\qquad
\lbrace\stackrel{a}{U},\stackrel{4}{V}\rbrace&=\left\lbrace
\begin{array}{ll}
\stackrel{a-4}{W}+\stackrel{a}{W}+\stackrel{a+4}{W}, & \parbox{
.2\linewidth}{$a\leq n-4$;}\\ \\
\stackrel{a-4}{W}+\stackrel{a}{W}, & \parbox{
.2\linewidth}{$a=n-3,\\ n-2$;}\\ \\
\stackrel{a-4}{W}, & \parbox{ .2\linewidth}{$a=n-1, n$.}
\end{array}
\right.
\end{flalign*}

For elements of small ranks we have
\begin{flalign*}
[\stackrel{1}{U},\stackrel{1}{V}]&=\left\lbrace
\begin{array}{ll}
\stackrel{2}{W}, & \mbox{\rm $n\geq 2$;}\\
0, & \mbox{\rm $n=1$.}
\end{array}
\right. &\qquad
\lbrace\stackrel{1}{U},\stackrel{1}{V}\rbrace&=\stackrel{0}{W}.
\\
[\stackrel{2}{U},\stackrel{1}{V}]&=\stackrel{1}{W}. &\qquad
\lbrace\stackrel{2}{U},\stackrel{1}{V}\rbrace&=\left\lbrace
\begin{array}{ll}
\stackrel{3}{W}, & \mbox{\rm $n\geq 2$;}\\
0, & \mbox{\rm $n=2$.}
\end{array}
\right.
\\
[\stackrel{2}{U},\stackrel{2}{V}]&=\left\lbrace
\begin{array}{ll}
\stackrel{2}{W}, & \mbox{\rm $n\geq 3$;}\\
0, & \mbox{\rm $n=2$.}
\end{array}
\right. &\qquad
\lbrace\stackrel{2}{U},\stackrel{2}{V}\rbrace&=\left\lbrace
\begin{array}{ll}
\stackrel{0}{W}+\stackrel{4}{W}, & \mbox{\rm $n\neq 2, 3$;}\\
\stackrel{0}{W}, & \mbox{\rm $n=2, 3$.}
\end{array}
\right.
\\
[\stackrel{3}{U},\stackrel{1}{V}]&=\left\lbrace
\begin{array}{ll}
\stackrel{4}{W}, & \mbox{\rm $n\geq 4$;}\\
0, & \mbox{\rm $n=3$.}
\end{array}
\right. &\qquad
\lbrace\stackrel{3}{U},\stackrel{1}{V}\rbrace&=\stackrel{2}{W}.
\\
[\stackrel{3}{U},\stackrel{2}{V}]&=\left\lbrace
\begin{array}{ll}
\stackrel{3}{W}, & \mbox{\rm $n\geq 4$;}\\
0, & \mbox{\rm $n=3$.}
\end{array}
\right. &\qquad
\lbrace\stackrel{3}{U},\stackrel{2}{V}\rbrace&=\left\lbrace
\begin{array}{ll}
\stackrel{1}{W}+\stackrel{5}{W}, & \mbox{\rm $n\neq 3, 4$;}\\
\stackrel{1}{W}, & \mbox{\rm $n=3, 4$.}
\end{array}
\right.
\\
[\stackrel{3}{U},\stackrel{3}{V}]&=\left\lbrace
\begin{array}{ll}
\stackrel{2}{W}+\stackrel{6}{W}, & \mbox{\rm $n\geq 6$;}\\
\stackrel{2}{W}, & \mbox{\rm $n=4,5$;}\\
0, & \mbox{\rm $n=3$.}
\end{array}
\right. &\qquad
\lbrace\stackrel{3}{U},\stackrel{3}{V}\rbrace&=\left\lbrace
\begin{array}{ll}
\stackrel{0}{W}+\stackrel{4}{W}, & \mbox{\rm $n\geq 5$;}\\
\stackrel{0}{W}, & \mbox{\rm $n=3,4$.}
\end{array}
\right.
\\
[\stackrel{4}{U},\stackrel{1}{V}]&=\stackrel{3}{W}. &\qquad
\lbrace\stackrel{4}{U},\stackrel{1}{V}\rbrace&=\left\lbrace
\begin{array}{ll}
\stackrel{5}{W}, & \mbox{\rm $n\geq 5$;}\\
0, & \mbox{\rm $n=4$.}
\end{array}
\right.
\\
[\stackrel{4}{U},\stackrel{2}{V}]&=\left\lbrace
\begin{array}{ll}
\stackrel{4}{W}, & \mbox{\rm $n\geq 5$;}\\
0, & \mbox{\rm $n=4$.}
\end{array}
\right. &\qquad
\lbrace\stackrel{4}{U},\stackrel{2}{V}\rbrace&=\left\lbrace
\begin{array}{ll}
\stackrel{2}{W}+\stackrel{6}{W}, & \mbox{\rm $n\neq 4, 5$;}\\
\stackrel{2}{W}, & \mbox{\rm $n=4, 5$.}
\end{array}
\right.
\\
[\stackrel{4}{U},\stackrel{3}{V}]&=\left\lbrace
\begin{array}{ll}
\stackrel{1}{W}+\stackrel{5}{W}, & \mbox{\rm $n\geq 6$;}\\
\stackrel{1}{W}, & \mbox{\rm $n=4, 5$.}
\end{array}
\right. &\qquad
\lbrace\stackrel{4}{U},\stackrel{3}{V}\rbrace&=\left\lbrace
\begin{array}{ll}
\stackrel{3}{W}+\stackrel{7}{W}, & \mbox{\rm $n\geq 7$;}\\
\stackrel{3}{W}, & \mbox{\rm $n=5,6$;}\\
0, & \mbox{\rm $n=4$.}
\end{array}
\right.
\\
[\stackrel{4}{U},\stackrel{4}{V}]&=\left\lbrace
\begin{array}{ll}
\stackrel{2}{W}+\stackrel{6}{W}, & \mbox{\rm $n\geq 7$;}\\
\stackrel{2}{W}, & \mbox{\rm $n=5, 6$;}\\
0, & \mbox{\rm $n=4$.}
\end{array}
\right. &\qquad
\lbrace\stackrel{4}{U},\stackrel{4}{V}\rbrace&=\left\lbrace
\begin{array}{ll}
\stackrel{0}{W}+\stackrel{4}{W}+\stackrel{8}{W}, & \mbox{\rm $n\geq 8$;}\\
\stackrel{0}{W}+\stackrel{4}{W}, & \mbox{\rm $n=6, 7$;}\\
0, & \mbox{\rm $n=4, 5$.}
\end{array}
\right.
\end{flalign*}

For elements of ranks that are closed to $n$ we have
\begin{flalign*}
[\stackrel{n}{U},\stackrel{n}{V}]&=0. &\qquad
\lbrace\stackrel{n}{U},\stackrel{n}{V}\rbrace&=\stackrel{0}{W}.
\\
[\stackrel{n}{U},\stackrel{n-1}{V}]&=\left\lbrace
\begin{array}{ll}
0, & \mbox{\rm n - odd;}\\
\stackrel{1}{W}, & \mbox{\rm n - even.}
\end{array}
\right. &\qquad
\lbrace\stackrel{n}{U},\stackrel{n-1}{V}\rbrace&=\left\lbrace
\begin{array}{ll}
0, & \mbox{\rm n - even;}\\
\stackrel{1}{W}, & \mbox{\rm n - odd.}
\end{array}
\right.
\\
[\stackrel{n}{U},\stackrel{n-2}{V}]&=0. &\qquad
\lbrace\stackrel{n}{U},\stackrel{n-2}{V}\rbrace&=\stackrel{2}{W}.
\\
[\stackrel{n-1}{U},\stackrel{n-1}{V}]&=\left\lbrace
\begin{array}{ll}
\stackrel{2}{W}, & \mbox{\rm $n\neq 1$;}\\
0, & \mbox{\rm $n=1$.}
\end{array}
\right. &\qquad
\lbrace\stackrel{n-1}{U},\stackrel{n-1}{V}\rbrace&=\stackrel{0}{W}.
\\
[\stackrel{n-1}{U},\stackrel{n-2}{V}]&=\left\lbrace
\begin{array}{ll}
\stackrel{1}{W}, & \mbox{\rm n - odd;}\\
\stackrel{3}{W}, & \mbox{\rm n - even, $n\neq 2$;}\\
0, & \mbox{\rm $n=2$.}
\end{array}
\right. &\qquad
\lbrace\stackrel{n-1}{U},\stackrel{n-2}{V}\rbrace&=\left\lbrace
\begin{array}{ll}
\stackrel{1}{W}, & \mbox{\rm n - even;}\\
\stackrel{3}{W}, & \mbox{\rm n - odd;.}
\end{array}
\right.
\\
[\stackrel{n-2}{U},\stackrel{n-2}{V}]&=\left\lbrace
\begin{array}{ll}
\stackrel{2}{W}, & \mbox{\rm $n\geq 3$;}\\
0, & \mbox{\rm $n=2$.}
\end{array}
\right. &\qquad
\lbrace\stackrel{n-2}{U},\stackrel{n-2}{V}\rbrace&=\left\lbrace
\begin{array}{ll}
\stackrel{0}{W}+\stackrel{4}{W}, & \mbox{\rm $n\geq 4$;}\\
\stackrel{0}{W}, & \mbox{\rm $n=2, 3$.}
\end{array}
\right.
\end{flalign*}

If rank of the second element is small and the rank of the first element is closed to $n$, then
\begin{flalign*}
[\stackrel{n}{U},\stackrel{1}{V}]&=\left\lbrace
\begin{array}{ll}
\stackrel{n-1}{W}, & \mbox{\rm n - even;}\\
0, & \mbox{\rm n - odd.}
\end{array}
\right. &\qquad
\lbrace\stackrel{n}{U},\stackrel{1}{V}\rbrace&=\left\lbrace
\begin{array}{ll}
0, & \mbox{\rm n - even;}\\
\stackrel{n-1}{W}, & \mbox{\rm n - odd.}
\end{array}
\right.
\\
[\stackrel{n-1}{U},\stackrel{1}{V}]&=\left\lbrace
\begin{array}{ll}
\stackrel{n}{W}, & \mbox{\rm n - even;}\\
\stackrel{n-2}{W}, & \mbox{\rm n - odd.}
\end{array}
\right. &\qquad
\lbrace\stackrel{n-1}{U},\stackrel{1}{V}\rbrace&=\left\lbrace
\begin{array}{ll}
\stackrel{n-2}{W}, & \mbox{\rm n - even;}\\
\stackrel{n}{W}, & \mbox{\rm n - odd.}
\end{array}
\right.
\\
[\stackrel{n-2}{U},\stackrel{1}{V}]&=\left\lbrace
\begin{array}{ll}
\stackrel{n-3}{W}, & \mbox{\rm n - even;}\\
\stackrel{n-1}{W}, & \mbox{\rm n - odd.}
\end{array}
\right. &\qquad
\lbrace\stackrel{n-2}{U},\stackrel{1}{V}\rbrace&=\left\lbrace
\begin{array}{ll}
\stackrel{n-1}{W}, & \mbox{\rm n - even;}\\
\stackrel{n-3}{W}, & \mbox{\rm n - odd.}
\end{array}
\right.
\\
[\stackrel{n}{U},\stackrel{2}{V}]&=0. &\qquad
\lbrace\stackrel{n}{U},\stackrel{2}{V}\rbrace&=\stackrel{n-2}{W}.
\\
[\stackrel{n-1}{U},\stackrel{2}{V}]&=\stackrel{n-1}{W}. &\qquad
\lbrace\stackrel{n-1}{U},\stackrel{2}{V}\rbrace&=\stackrel{n-3}{W}.
\\
[\stackrel{n-2}{U},\stackrel{2}{V}]&=\stackrel{n-2}{W}. &\qquad
\lbrace\stackrel{n-2}{U},\stackrel{2}{V}\rbrace&=\stackrel{n-4}{W}+\stackrel{n}{W}.
\end{flalign*}

The following tables illustrate formulas for commutators from Theorem 1.

For the dimensions $n=1, 2, \ldots 10$ of Clifford algebra we have

\begin{tabular}{|c||c|}
\hline
n=1 & 1 \\ \hline \hline
1 & - \\ \hline
\end{tabular} \quad
\begin{tabular}{|c||c|c|}
\hline
n=2 & 1 & 2 \\ \hline \hline
1 & 2 & 1 \\ \hline
2 & 1 & -\\ \hline
\end{tabular} \quad
\begin{tabular}{|c||c|c|c|}
\hline
n=3 & 1 & 2 & 3 \\ \hline \hline
1 & 2 & 1 & - \\ \hline
2 & 1 & 2 & - \\ \hline
3 & - & - & - \\ \hline
\end{tabular} \quad
\begin{tabular}{|c||c|c|c|c|}
\hline
n=4 & 1 & 2 & 3 & 4 \\ \hline \hline
1 & 2 & 1 & 4 & 3 \\ \hline
2 & 1 & 2 & 3 & - \\ \hline
3 & 4 & 3 & 2 & 1 \\ \hline
4 & 3 & - & 1 & - \\ \hline
\end{tabular}

\bigskip
\begin{tabular}{|c||c|c|c|c|c|}
\hline
n=5 & 1 & 2 & 3 & 4 & 5 \\ \hline \hline
1 & 2 & 1 & 4 & 3 & - \\ \hline
2 & 1 & 2 & 3 & 4 & - \\ \hline
3 & 4 & 3 & 2 & 1 & - \\ \hline
4 & 3 & 4 & 1 & 2 & - \\ \hline
5 & - & - & - & - & - \\ \hline
\end{tabular} \quad
\begin{tabular}{|c||c|c|c|c|c|c|}
\hline
n=6 & 1 & 2 & 3 & 4 & 5 & 6 \\ \hline \hline
1 & 2 & 1 & 4 & 3 & 6 & 5 \\ \hline
2 & 1 & 2 & 3 & 4 & 5 & - \\ \hline
3 & 4 & 3 & 2/6 & 1/5 & 4 & 3 \\ \hline
4 & 3 & 4 & 1/5 & 2 & 3 & - \\ \hline
5 & 6 & 5 & 4 & 3 & 2 & 1 \\ \hline
6 & 5 & - & 3 & - & 1 & - \\ \hline
\end{tabular}

\bigskip
\begin{tabular}{|c||c|c|c|c|c|c|c|}
\hline
n=7 & 1 & 2 & 3 & 4 & 5 & 6 & 7 \\ \hline \hline
1 & 2 & 1 & 4 & 3 & 6 & 5 & - \\ \hline
2 & 1 & 2 & 3 & 4 & 5 & 6 & - \\ \hline
3 & 4 & 3 & 2/6 & 1/5 & 4 & 3 & - \\ \hline
4 & 3 & 4 & 1/5 & 2/6 & 3 & 4 & - \\ \hline
5 & 6 & 5 & 4 & 3 & 2 & 1 & - \\ \hline
6 & 5 & 6 & 3 & 4 & 1 & 2 & - \\ \hline
7 & - & - & - & - & - & - & - \\ \hline
\end{tabular}

\bigskip
\begin{tabular}{|c||c|c|c|c|c|c|c|c|}
\hline
n=8 & 1 & 2 & 3 & 4 & 5 & 6 & 7 & 8 \\ \hline \hline
1 & 2 & 1 & 4 & 3 & 6 & 5 & 8 & 7 \\ \hline
2 & 1 & 2 & 3 & 4 & 5 & 6 & 7 & - \\ \hline
3 & 4 & 3 & 2/6 & 1/5 & 4/8 & 3/7 & 6 & 5 \\ \hline
4 & 3 & 4 & 1/5 & 2/6 & 3/7 & 4 & 5 & - \\ \hline
5 & 6 & 5 & 4/8 & 3/7 & 2/6 & 1/5 & 4 & 3 \\ \hline
6 & 5 & 6 & 3/7 & 4 & 1/5 & 2 & 3 & - \\ \hline
7 & 8 & 7 & 6 & 5 & 4 & 3 & 2 & 1 \\ \hline
8 & 7 & - & 5 & - & 3 & - & 1 & - \\ \hline
\end{tabular}

\bigskip
\begin{tabular}{|c||c|c|c|c|c|c|c|c|c|}
\hline
n=9 & 1 & 2 & 3 & 4 & 5 & 6 & 7 & 8 & 9 \\ \hline \hline
1 & 2 & 1 & 4 & 3 & 6 & 5 & 8 & 7 & - \\ \hline
2 & 1 & 2 & 3 & 4 & 5 & 6 & 7 & 8 & - \\ \hline
3 & 4 & 3 & 2/6 & 1/5 & 4/8 & 3/7 & 6 & 5 & - \\ \hline
4 & 3 & 4 & 1/5 & 2/6 & 3/7 & 4/8 & 5 & 6 & - \\ \hline
5 & 6 & 5 & 4/8 & 3/7 & 2/6 & 1/5 & 4 & 3 & - \\ \hline
6 & 5 & 6 & 3/7 & 4/8 & 1/5 & 2/6 & 3 & 4 & - \\ \hline
7 & 8 & 7 & 6 & 5 & 4 & 3 & 2 & 1 & - \\ \hline
8 & 7 & 8 & 5 & 6 & 3 & 4 & 1 & 2 & - \\ \hline
9 & - & - & - & - & - & - & - & - & - \\ \hline
\end{tabular}

\bigskip
\begin{tabular}{|c||c|c|c|c|c|c|c|c|c|c|}
\hline
n=10 & 1 & 2 & 3 & 4 & 5 & 6 & 7 & 8 & 9 & 10 \\ \hline \hline
1 & 2 & 1 & 4 & 3 & 6 & 5 & 8 & 7 & 10 & 9\\ \hline
2 & 1 & 2 & 3 & 4 & 5 & 6 & 7 & 8 & 9 & - \\ \hline
3 & 4 & 3 & 2/6 & 1/5 & 4/8 & 3/7 & 6/10 & 5/9 & 8 & 7 \\ \hline
4 & 3 & 4 & 1/5 & 2/6 & 3/7 & 4/8 & 5/9 & 6 & 7 & - \\ \hline
5 & 6 & 5 & 4/8 & 3/7 & 2/6/10 & 1/5/9 & 4/8 & 3/7 & 6 & 5 \\ \hline
6 & 5 & 6 & 3/7 & 4/8 & 1/5/9 & 2/6 & 3/7 & 4 & 5 & - \\ \hline
7 & 8 & 7 & 6/10 & 5/9 & 4/8 & 3/7 & 2/6 & 1/5 & 4 & 3 \\ \hline
8 & 7 & 8 & 5/9 & 6 & 3/7 & 4 & 1/5 & 2 & 3 & - \\ \hline
9 & 10 & 9 & 8 & 7 & 6 & 5 & 4 & 3 & 2 & 1 \\ \hline
10 & 9 & - & 7 & - & 5 & - & 3 & - & 1 & - \\ \hline
\end{tabular}

\bigskip
Tables are symmetric with respect to the main diagonal because
$$
[\stackrel{k}{U}, \stackrel{l}{V}] = - [\stackrel{l}{V},
\stackrel{k}{U}].
$$

For anticommutators we have the following tables ($n=1, 2, \ldots 10$):

\begin{tabular}{|c||c|}
\hline
n=1 & 1 \\ \hline \hline
1 & 0 \\ \hline
\end{tabular} \quad
\begin{tabular}{|c||c|c|}
\hline
n=2 & 1 & 2 \\ \hline \hline
1 & 0 & - \\ \hline
2 & - & 0\\ \hline
\end{tabular} \quad
\begin{tabular}{|c||c|c|c|}
\hline
n=3 & 1 & 2 & 3 \\ \hline \hline
1 & 0 & 3 & 2 \\ \hline
2 & 3 & 0 & 1 \\ \hline
3 & 2 & 1 & 0 \\ \hline
\end{tabular} \quad
\begin{tabular}{|c||c|c|c|c|}
\hline
n=4 & 1 & 2 & 3 & 4 \\ \hline \hline
1 & 0 & 3 & 2 & - \\ \hline
2 & 3 & 0/4 & 1 & 2 \\ \hline
3 & 2 & 1 & 0 & - \\ \hline
4 & - & 2 & - & 0 \\ \hline
\end{tabular}

\bigskip
\begin{tabular}{|c||c|c|c|c|c|}
\hline
n=5 & 1 & 2 & 3 & 4 & 5 \\ \hline \hline
1 & 0 & 3 & 2 & 5 & 4 \\ \hline
2 & 3 & 0/4 & 1/5 & 2 & 3 \\ \hline
3 & 2 & 1/5 & 0/4 & 3 & 2 \\ \hline
4 & 5 & 2 & 3 & 0 & 1 \\ \hline
5 & 4 & 3 & 2 & 1 & 0 \\ \hline
\end{tabular} \quad
\begin{tabular}{|c||c|c|c|c|c|c|}
\hline
n=6 & 1 & 2 & 3 & 4 & 5 & 6 \\ \hline \hline
1 & 0 & 3 & 2 & 5 & 4 & - \\ \hline
2 & 3 & 0/4 & 1/5 & 2/6 & 3 & 4 \\ \hline
3 & 2 & 1/5 & 0/4 & 3 & 2 & - \\ \hline
4 & 5 & 2/6 & 3 & 0/4 & 1 & 2 \\ \hline
5 & 4 & 3 & 2 & 1 & 0 & - \\ \hline
6 & - & 4 & - & 2 & - & 0 \\ \hline
\end{tabular}

\bigskip
\begin{tabular}{|c||c|c|c|c|c|c|c|}
\hline
n=7 & 1 & 2 & 3 & 4 & 5 & 6 & 7 \\ \hline \hline
1 & 0 & 3 & 2 & 5 & 4 & 7 & 6 \\ \hline
2 & 3 & 0/4 & 1/5 & 2/6 & 3/7 & 4 & 5 \\ \hline
3 & 2 & 1/5 & 0/4 & 13/7 & 2/6 & 5 & 4 \\ \hline
4 & 5 & 2/6 & 3/7 & 0/4 & 1/5 & 2 & 3 \\ \hline
5 & 4 & 3/7 & 2/6 & 1/5 & 0/4 & 3 & 2 \\ \hline
6 & 7 & 4 & 5 & 2 & 3 & 0 & 1 \\ \hline
7 & 6 & 5 & 4 & 3 & 2 & 1 & 0 \\ \hline
\end{tabular}

\bigskip
\begin{tabular}{|c||c|c|c|c|c|c|c|c|}
\hline
n=8 & 1 & 2 & 3 & 4 & 5 & 6 & 7 & 8 \\ \hline \hline
1 & 0 & 3 & 2 & 5 & 4 & 7 & 6 & - \\ \hline
2 & 3 & 0/4 & 1/5 & 2/6 & 3/7 & 4/8 & 5 & 6 \\ \hline
3 & 2 & 1/5 & 0/4 & 3/7 & 2/6 & 5 & 4 & - \\ \hline
4 & 5 & 2/6 & 3/7 & 0/4/8 & 1/5 & 2/6 & 3 & 4 \\ \hline
5 & 4 & 3/7 & 2/6 & 1/5 & 0/4 & 3 & 2 & - \\ \hline
6 & 7 & 4/8 & 5 & 2/6 & 3 & 0/4 & 1 & 2 \\ \hline
7 & 6 & 5 & 4 & 3 & 2 & 1 & 0 & - \\ \hline
8 & - & 6 & - & 4 & - & 2 & - & 0 \\ \hline
\end{tabular}

\bigskip
\begin{tabular}{|c||c|c|c|c|c|c|c|c|c|}
\hline
n=9 & 1 & 2 & 3 & 4 & 5 & 6 & 7 & 8 & 9 \\ \hline \hline
1 & 0 & 3 & 2 & 5 & 4 & 7 & 6 & 9 & 8\\ \hline
2 & 3 & 0/4 & 1/5 & 2/6 & 3/7 & 4/8 & 5/9 & 6 & 7 \\ \hline
3 & 2 & 1/5 & 0/4 & 3/7 & 2/6 & 5/9 & 4/8 & 7 & 6 \\ \hline
4 & 5 & 2/6 & 3/7 & 0/4/8 & 1/5/9 & 2/6 & 3/7 & 4 & 5 \\ \hline
5 & 4 & 3/7 & 2/6 & 1/5/9 & 0/4/8 & 3/7 & 2/6 & 5 & 4 \\ \hline
6 & 7 & 4/8 & 5/9 & 2/6 & 3/7 & 0/4 & 1/5 & 2 & 3 \\ \hline
7 & 6 & 5/9 & 4/8 & 3/7 & 2/6 & 1/5 & 0/4 & 3 & 2 \\ \hline
8 & 9 & 6 & 7 & 4 & 5 & 2 & 3 & 0 & 1 \\ \hline
9 & 8 & 7 & 6 & 5 & 4 & 3 & 2 & 1 & 0 \\ \hline
\end{tabular}

\bigskip
\begin{tabular}{|c||c|c|c|c|c|c|c|c|c|c|}
\hline
n=10 & 1 & 2 & 3 & 4 & 5 & 6 & 7 & 8 & 9 & 10 \\ \hline \hline
1 & 0 & 3 & 2 & 5 & 4 & 7 & 6 & 9 & 8 & -\\ \hline
2 & 3 & 0/4 & 1/5 & 2/6 & 3/7 & 4/8 & 5/9 & 6/10 & 7 & 8 \\ \hline
3 & 2 & 1/5 & 0/4 & 3/7 & 2/6 & 5/9 & 4/8 & 7 & 6 & - \\ \hline
4 & 5 & 2/6 & 3/7 & 0/4/8 & 1/5/9 & 2/6/10 & 3/7 & 4/8 & 5 & 6 \\ \hline
5 & 4 & 3/7 & 2/6 & 1/5/9 & 0/4/8 & 3/7 & 2/6 & 5 & 4 & - \\ \hline
6 & 7 & 4/8 & 5/9 & 2/6/10 & 3/7 & 0/4/8 & 1/5 & 2/6 & 3 & 4 \\ \hline
7 & 6 & 5/9 & 4/8 & 3/7 & 2/6 & 1/5 & 0/4 & 3 & 2 & - \\ \hline
8 & 9 & 6/10 & 7 & 4/8 & 5 & 2/6 & 3 & 0/4 & 1 & 2 \\ \hline
9 & 8 & 7 & 6 & 5 & 4 & 3 & 2 & 1 & 0 & - \\ \hline
10 & - & 8 & - & 6 & - & 4 & - & 2 & - & 0 \\ \hline
\end{tabular}


\section{Subalgebras of the Lie algebra of pseudounitary group}
Consider the following set of Clifford algebra elements
$$
\Wcl(p,q)=\{U\in\cl(p,q): U^* U=e\},
$$
where * is the operation of Clifford conjugation
\cite{Marchuk:Shirokov} with properties
$$
e^*=e,\quad (e^{a_1}e^{a_2}\ldots e^{a_k})^*=e^{a_k}\ldots e^{a_1},\quad
(\lambda)^*=\overline{\lambda},
$$
$\lambda$ is a complex number and $\overline{\lambda}$ is the
conjugated complex number. This set forms a (Lie) group with respect
to the Clifford product and this group is called {\it the
pseudounitary group of Clifford algebra $\cl(p,q)$ }.

The set of elements with the commutator $[U,V]=UV-VU$
$$
\wcl(p,q)=\{u\in\cl(p,q): u^*=-u\}
$$
is {\it the Lie algebra of the Lie group $\Wcl(p,q)$}.

From this definition and from the definition of Clifford conjugation  it follows that an arbitrary element of this Lie algebra has the form
$$
u=i\stackrel{0}{u}+i\stackrel{1}{u}+\stackrel{2}{u}+\stackrel{3}{u}+i\stackrel{4}{u}+i\stackrel{5}{u}+\ldots+
a_n\stackrel{n}{u},
$$
i.e.
$$
u=\sum_{k=0}^{n}a_k \stackrel{k}{u},
$$
where
$$
a_k=\left\lbrace
\begin{array}{ll}
1, & \mbox{\rm $k=2, 3, 6, 7, \ldots$;}\\
i, & \mbox{\rm $k=0, 1, 4, 5, \ldots$,}
\end{array}
\right.
$$
and $\stackrel{k}{u} \in\cl_k^\R (p,q)$. By $\cl^\R (p,q)$  we
denote the real Clifford algebra.

We want to find direct sums of vector spaces $a_k\cl_k^\R(p,q)$ such that they form a Lie algebra (closed with respect to the commutator).
\begin{theorem}3.
Let $p+q=n$. The following 12 types of direct sums of vector spaces
$\stackrel{k}{u}$ are closed with respect to the commutator and,
hence, form subalgebras of Lie algebra $\wcl(p,q)$:
\begin{description}
\item 1) for $n\geq 1$: $$i\stackrel{0}{u};$$
\item 2) for $n\geq 1$: $$a_{n}\stackrel{n}{u};$$
\item 3) for $n\geq 2$: $$i\stackrel{1}{u}+\stackrel{2}{u};$$
\item 4) for $n\geq 3$ (if $n=2$ it is the same as item 2): $$\stackrel{2}{u};$$
\item 5) for $n\geq 4$ (if $n=2, 3$ it is the same as item 3): $$i\stackrel{1}{u}+\stackrel{2}{u}+\ldots+a_{n}\stackrel{n}{u}$$ for even $n$, $$i\stackrel{1}{u}+\stackrel{2}{u}+\ldots+a_{n-1}\stackrel{n-1}{u}$$ for odd $n$;
\item 6) for $n\geq 4$: $$\stackrel{2}{u}+a_{n-1}\stackrel{n-1}{u};$$
\item 7) for $n\geq 5$: $$\stackrel{2}{u}+a_{n-2}\stackrel{n-2}{u};$$
\item 8) for $n\geq 6$ (if $n=5$ it is the same as item 5): $$i\stackrel{1}{u}+\stackrel{2}{u}+a_{n-2}\stackrel{n-2}{u}+a_{n-1}\stackrel{n-1}{u}$$ for odd $n$ , $$i\stackrel{1}{u}+\stackrel{2}{u}+a_{n-1}\stackrel{n-1}{u}+a_{n}\stackrel{n}{u}$$ for even $n$;
\item 9) for $n\geq 6$ (if $n=2, 3$ it is the same as item 4, if $n=4$ it is the same as item 6, if $n=5$ it is the same as item 7):$$\stackrel{2}{u}+\stackrel{3}{u}+\stackrel{6}{u}+\stackrel{7}{u}+\stackrel{10}{u}+\stackrel{11}{u}+\ldots+
\stackrel{k}{u}$$ for $n=k+1, k+2$ for odd $k$ and $n=k, k+1$ for
even $k$;
\item 10) for $n\geq 7$ (if $n=3, 4$ it is the same as item 4, if $n=5$ it is the same as item 6, if $n=6$ it is the same as item 7): $$\stackrel{2}{u}+i\stackrel{4}{u}+\stackrel{6}{u}+i\stackrel{8}{u}+\stackrel{10}{u}+i\stackrel{12}{u}+\ldots+
a_{k}\stackrel{k}{u}$$ for $n=k+1, k+2$;
\item 11) for $n\geq 8$ (if $n=2, 3, 4, 5$ it is the same as item 3, if $n=6, 7$ it is the same as item 8): $$i\stackrel{1}{u}+\stackrel{2}{u}+i\stackrel{5}{u}+\stackrel{6}{u}+i\stackrel{9}{u}+\stackrel{10}{u}+\ldots+
a_{k}\stackrel{k}{u}$$ for $n=k, k+1, k+2, k+3$ for even $k$;
\item 12) for $n\geq 9$ (if $n=3, 4, 5, 6$ it is the same as item 4, if $n=7$ it is the same as item 6, if $n=8$ it is the same as item 7): $$\stackrel{2}{u}+\stackrel{6}{u}+\stackrel{10}{u}+\stackrel{14}{u}+\stackrel{18}{u}+\stackrel{22}{u}+\ldots+\stackrel{k}{u}$$ for $n=k+1, k+2, k+3, k+4$.
\end{description}
We can add $i\stackrel{0}{u}$ to any of these subalgebras. We can
add $a_n\stackrel{n}{u}$ to all types of subalgebras for odd $n$.
Also we can add $a_n\stackrel{n}{u}$ to subalgebras that consist of
elements of even ranks for even $n$. (In these cases we get
reducible subalgebras.)

\end{theorem}

\proof. \, Denote by $\sum_{j=b_1}^{b_k} \stackrel{j}{u}$ the
arbitrary element of the subspace $\cl_{b_1}(p,q) \oplus
\cl_{b_2}(p,q) \oplus \ldots  \oplus \cl_{b_k}(p,q)$. This subspace
form a subalgebra if $[\sum_{j=b_1}^{b_k}\stackrel{j}{u},
\sum_{j=b_1}^{b_k}\stackrel{j}{v}]$ can be written as
$\sum_{j=b_1}^{b_k}\stackrel{j}{w}$. That means
$[\stackrel{s}{u},\stackrel{t}{v}]$ can be written as
$\sum_{j=b_1}^{b_k}\stackrel{j}{w}$ for all $s,t = b_1, b_2, \ldots,
b_k$.

With the aid of Theorem 1 the proof of this theorem is
straightforward. \fin

We have the analogous theorem for Lie subalgebras of real or complex Clifford algebra:
\begin{theorem}4.
Consider the Clifford algebra $\cl(p,q)$ as an Lie algebra closed
with respect to the commutator $[U,V]=UV-VU$. Then the following 12
types of subspaces form Lie subalgebras of real (or complex)
Clifford algebra $\cl(p,q)$:
\begin{description}
\item 1) for $n\geq 1$: $$\stackrel{0}{u};$$
\item 2) for $n\geq 1$: $$\stackrel{n}{u};$$
\item 3) for $n\geq 2$: $$\stackrel{1}{u}+\stackrel{2}{u};$$
\item 4) for $n\geq 3$ (if $n=2$ it is the same as item 2): $$\stackrel{2}{u};$$
\item 5) for $n\geq 4$ (if $n=2, 3$ it is the same as item 3): $$\stackrel{1}{u}+\stackrel{2}{u}+\ldots+\stackrel{n}{u}$$ for even $n$, $$\stackrel{1}{u}+\stackrel{2}{u}+\ldots+\stackrel{n-1}{u}$$ for odd $n$;
\item 6) for $n\geq 4$: $$\stackrel{2}{u}+\stackrel{n-1}{u};$$
\item 7) for $n\geq 5$: $$\stackrel{2}{u}+\stackrel{n-2}{u};$$
\item 8) for $n\geq 6$ (if $n=5$ it is the same as item 5): $$\stackrel{1}{u}+\stackrel{2}{u}+\stackrel{n-2}{u}+\stackrel{n-1}{u}$$ for odd $n$ , $$\stackrel{1}{u}+\stackrel{2}{u}+\stackrel{n-1}{u}+\stackrel{n}{u}$$ for even $n$;
\item 9) for $n\geq 6$ (if $n=2, 3$ it is the same as item 4, if $n=4$ it is the same as item 6, if $n=5$ it is the same as item 7):$$\stackrel{2}{u}+\stackrel{3}{u}+\stackrel{6}{u}+\stackrel{7}{u}+\stackrel{10}{u}+\stackrel{11}{u}+\ldots+
\stackrel{k}{u}$$ for $n=k+1, k+2$ for odd $k$ and $n=k, k+1$ for
even $k$;
\item 10) for $n\geq 7$ (if $n=3, 4$ it is the same as item 4, if $n=5$ it is the same as item 6, if $n=6$ it is the same as item 7): $$\stackrel{2}{u}+\stackrel{4}{u}+\stackrel{6}{u}+\stackrel{8}{u}+\stackrel{10}{u}+\stackrel{12}{u}+\ldots+\stackrel{k}{u}$$ for $n=k+1, k+2$;
\item 11) for $n\geq 8$ (if $n=2, 3, 4, 5$ it is the same as item 3, if $n=6, 7$ it is the same as item 8): $$\stackrel{1}{u}+\stackrel{2}{u}+\stackrel{5}{u}+\stackrel{6}{u}+\stackrel{9}{u}+\stackrel{10}{u}+\ldots+\stackrel{k}{u}$$ for $n=k, k+1, k+2, k+3$ for even $k$;
\item 12) for $n\geq 9$ (if $n=3, 4, 5, 6$ it is the same as item 4, if $n=7$ it is the same as item 6, if $n=8$ it is the same as item 7): $$\stackrel{2}{u}+\stackrel{6}{u}+\stackrel{10}{u}+\stackrel{14}{u}+\stackrel{18}{u}+\stackrel{22}{u}+\ldots+\stackrel{k}{u}$$ for $n=k+1, k+2, k+3, k+4$.
\end{description}
We can add $\stackrel{0}{u}$ to any of these subalgebras. We can add
$\stackrel{n}{u}$ to all types of subalgebras for odd $n$. Also we
can add $\stackrel{n}{u}$ to subalgebras that consist of elements of
even ranks for even $n$. (In these cases we get reducible
subalgebras.)

\end{theorem}

\proof. \, If we replace $a_k$ by $1$ in the proof of Theorem
3, we obtain the proof of this theorem. \fin

Let's write down all subalgebras of the Lie algebra of pseudounitary
group of Clifford algebra for the dimensions n=1, 2, \ldots 10. We
are interested only in Lie algebras that are direct sums of elements
of different ranks. If we replace $a_k$ by $1$, we will get Lie
subalgebras of Clifford algebra.

\bigskip
For n=1:

1) $i\stackrel{0}{u}$, 2) $i\stackrel{1}{u}$.

For n=2:

1) $i\stackrel{0}{u}$, 2) $\stackrel{2}{u}$, 3)
$i\stackrel{1}{u}+\stackrel{2}{u}$.

For n=3:

1) $i\stackrel{0}{u}$, 2) $\stackrel{3}{u}$, 3)
$i\stackrel{1}{u}+\stackrel{2}{u}$, 4) $\stackrel{2}{u}$.

For n=4:

1) $i\stackrel{0}{u}$, 2) $i\stackrel{4}{u}$, 3)
$i\stackrel{1}{u}+\stackrel{2}{u}$, 4) $\stackrel{2}{u}$, 5)
$i\stackrel{1}{u}+\stackrel{2}{u}+\stackrel{3}{u}+i\stackrel{4}{u}$,
6) $\stackrel{2}{u}+\stackrel{3}{u}$.

For n=5:

1) $i\stackrel{0}{u}$, 2) $i\stackrel{5}{u}$, 3)
$i\stackrel{1}{u}+\stackrel{2}{u}$, 4) $\stackrel{2}{u}$, 5)
$i\stackrel{1}{u}+\stackrel{2}{u}+\stackrel{3}{u}+i\stackrel{4}{u}$,
6) $\stackrel{2}{u}+i\stackrel{4}{u}$, 7)
$\stackrel{2}{u}+\stackrel{3}{u}$.

For n=6:

1) $i\stackrel{0}{u}$, 2) $\stackrel{6}{u}$, 3)
$i\stackrel{1}{u}+\stackrel{2}{u}$, 4) $\stackrel{2}{u}$, 5)
$i\stackrel{1}{u}+\stackrel{2}{u}+\stackrel{3}{u}+i\stackrel{4}{u}+i\stackrel{5}{u}+\stackrel{6}{u}$,
6) $\stackrel{2}{u}+i\stackrel{5}{u}$, 7)
$\stackrel{2}{u}+i\stackrel{4}{u}$, 8)
$i\stackrel{1}{u}+\stackrel{2}{u}+i\stackrel{5}{u}+\stackrel{6}{u}$,
9) $\stackrel{2}{u}+\stackrel{3}{u}+\stackrel{6}{u}$.

For n=7:

1) $i\stackrel{0}{u}$, 2) $\stackrel{7}{u}$, 3)
$i\stackrel{1}{u}+\stackrel{2}{u}$, 4) $\stackrel{2}{u}$, 5)
$i\stackrel{1}{u}+\stackrel{2}{u}+\stackrel{3}{u}+i\stackrel{4}{u}+i\stackrel{5}{u}+\stackrel{6}{u}$,
6) $\stackrel{2}{u}+\stackrel{6}{u}$, 7)
$\stackrel{2}{u}+i\stackrel{5}{u}$, 8)
$i\stackrel{1}{u}+\stackrel{2}{u}+i\stackrel{5}{u}+\stackrel{6}{u}$,
9) $\stackrel{2}{u}+\stackrel{3}{u}+\stackrel{6}{u}$, 10)
$\stackrel{2}{u}+i\stackrel{4}{u}+\stackrel{6}{u}$.

For n=8:

1) $i\stackrel{0}{u}$, 2) $i\stackrel{8}{u}$, 3)
$i\stackrel{1}{u}+\stackrel{2}{u}$, 4) $\stackrel{2}{u}$, 5)
$i\stackrel{1}{u}+\stackrel{2}{u}+\stackrel{3}{u}+i\stackrel{4}{u}+i\stackrel{5}{u}+\stackrel{6}{u}+\stackrel{7}{u}+i\stackrel{8}{u}$,
6) $\stackrel{2}{u}+\stackrel{7}{u}$, 7)
$\stackrel{2}{u}+\stackrel{6}{u}$, 8)
$i\stackrel{1}{u}+\stackrel{2}{u}+\stackrel{7}{u}+i\stackrel{8}{u}$,
9)
$\stackrel{2}{u}+\stackrel{3}{u}+\stackrel{6}{u}+\stackrel{7}{u}$,
10) $\stackrel{2}{u}+i\stackrel{4}{u}+\stackrel{6}{u}$, 11)
$i\stackrel{1}{u}+\stackrel{2}{u}+i\stackrel{5}{u}+\stackrel{6}{u}$.

For n=9:

1) $i\stackrel{0}{u}$, 2) $i\stackrel{9}{u}$, 3)
$i\stackrel{1}{u}+\stackrel{2}{u}$, 4) $\stackrel{2}{u}$, 5)
$i\stackrel{1}{u}+\stackrel{2}{u}+\stackrel{3}{u}+i\stackrel{4}{u}+i\stackrel{5}{u}+\stackrel{6}{u}+\stackrel{7}{u}+i\stackrel{8}{u}$,
6) $\stackrel{2}{u}+i\stackrel{8}{u}$, 7)
$\stackrel{2}{u}+\stackrel{7}{u}$, 8)
$i\stackrel{1}{u}+\stackrel{2}{u}+\stackrel{7}{u}+i\stackrel{8}{u}$,
9)
$\stackrel{2}{u}+\stackrel{3}{u}+\stackrel{6}{u}+\stackrel{7}{u}$,
10)
$\stackrel{2}{u}+i\stackrel{4}{u}+\stackrel{6}{u}+i\stackrel{8}{u}$,
11)
$i\stackrel{1}{u}+\stackrel{2}{u}+i\stackrel{5}{u}+\stackrel{6}{u}$,
12) $\stackrel{2}{u}+\stackrel{6}{u}$.

For n=10:

1) $i\stackrel{0}{u}$, 2) $\stackrel{10}{u}$, 3)
$i\stackrel{1}{u}+\stackrel{2}{u}$, 4) $\stackrel{2}{u}$, 5)
$i\stackrel{1}{u}+\stackrel{2}{u}+\stackrel{3}{u}+i\stackrel{4}{u}+i\stackrel{5}{u}+\stackrel{6}{u}+\stackrel{7}{u}+i\stackrel{8}{u}+i\stackrel{9}{u}+\stackrel{10}{u}$,
6) $\stackrel{2}{u}+i\stackrel{9}{u}$, 7)
$\stackrel{2}{u}+i\stackrel{8}{u}$, 8)
$i\stackrel{1}{u}+\stackrel{2}{u}+i\stackrel{9}{u}+\stackrel{10}{u}$,
9)
$\stackrel{2}{u}+\stackrel{3}{u}+\stackrel{6}{u}+\stackrel{7}{u}+\stackrel{10}{u}$,
10)
$\stackrel{2}{u}+i\stackrel{4}{u}+\stackrel{6}{u}+i\stackrel{8}{u}$,
11)
$i\stackrel{1}{u}+\stackrel{2}{u}+i\stackrel{5}{u}+\stackrel{6}{u}+i\stackrel{9}{u}+\stackrel{10}{u}$,
12) $\stackrel{2}{u}+\stackrel{6}{u}$.

We can add $i\stackrel{0}{u}$ to any of these subalgebras. We can
add $a_n\stackrel{n}{u}$ to all types of subalgebras for odd $n$.
Also we can add $a_n\stackrel{n}{u}$ to subalgebras that consist of
elements of even ranks for even $n$. We have reducible subalgebras
in these cases.

\bigskip

\noindent{\bf Acknowledgements.} The author is grateful to
N.~G.~Marchuk for constant attention to this work.

\end{document}